# RuDaCoP: The Dataset for Smartphone-based Intellectual Pedestrian Navigation

Andrey Bayev
*Moscow Research Center*
*Huawei Technologies Co., Ltd.*
Moscow, Russia
bayev.andrey@huawei.com

Ivan Chistyakov
*Moscow Research Center*
*Huawei Technologies Co., Ltd.*
Moscow, Russia
chistyakov.ivan@huawei.com

Alexey Derevyankin
*Moscow Research Center*
*Huawei Technologies Co., Ltd.*
Moscow, Russia
derevyankin.alexey@huawei.com

Ilya Gartseev
*Moscow Research Center*
*Huawei Technologies Co., Ltd.*
Moscow, Russia
gartseev.ilia@huawei.com

Alexey Nikulin
*Moscow Research Center*
*Huawei Technologies Co., Ltd.*
Moscow, Russia
nikulin.alexey@huawei.com

Mikhail Pikhletsky
*Moscow Research Center*
*Huawei Technologies Co., Ltd.*
Moscow, Russia
pikhletsky.mikhail@huawei.com

*Abstract*—This paper presents the large and diverse dataset for development of smartphone-based pedestrian navigation algorithms. This dataset consists of about 1200 sets of inertial measurements from sensors of several smartphones. The measurements are collected while walking through different trajectories up to 10 minutes long. The data are accompanied by the high accuracy ground truth collected with two foot-mounted inertial measurement units and post-processed by the presented algorithms. The dataset suits both for training of intellectual pedestrian navigation algorithms based on learning techniques and for development of pedestrian navigation algorithms based on classical approaches. The dataset is accessible at http://gartseev.ru/projects/ipin2019.

## I. INTRODUCTION

Problems of pedestrian navigation and localization (PNL) fill an important place among all navigation problems nowadays. PNL-based systems are used by fire and rescue services for immediate orientation inside buildings in cases of emergencies such as explosions and fires. These systems help visitors to orient themselves in large buildings such as airports, railway stations, and shopping centers. This task is also a major track for visually impaired people.

A variety of information sources may be used for PNL problem solution: satellite navigation systems, Wi-Fi, Bluetooth, and other signals. However, it is not always possible to rely on these signals. Firstly, their application requires an appropriate infrastructure that is not always present. Secondly, even if the required infrastructure is available, outer signals may not have enough power and quality (for example, it is well known that GPS signals are hardly suitable for indoor navigation). That is why inertial sensors, which do not require any external information, are widely used in PNL problems. Accordingly, development of inertial navigation algorithms that consider specifics of PNL problems is a crucial task at present. Inertial navigation for the purpose of PNL is called pedestrian dead reckoning (PDR).

Primarily, PNL specific is in low accuracy of mass-producible sensors that are used in smartphones. In this case traditional inertial navigation algorithms based on a double integration of acceleration yield non-satisfactory results because positioning errors attain unacceptable values very quickly due to low sensors accuracy [1], [2]. This holds even if one calibrates sensors before navigation. Therefore, other algorithms are typically used for PDR problems: they are usually based on an estimation of step length and heading angle. The estimation is performed by accelerometers, gyroscopes, and magnetometers, which are present in most devices. Such algorithms yield a more suitable solution. Thus, developing the PDR algorithms, one faces with several sub-problems that are not typical for classical inertial navigation: step length estimation (SLE), determination of the moments when a foot touches the ground (or lifts off the ground), the periods when a foot is motionless, etc.

For smartphone-based PDR algorithms, there is an additional difficulty. A smartphone may be held in different places (for example, in a hand, in a pocket, or in a bag as it is shown in Fig. 1(a) with orange boxes) and, consequently, it may move irregularly with respect to its holder. One of the approaches to overcome this difficulty is usage of machine learning (ML) and, particularly, deep learning methods while solving the above listed sub-problems [3], [4]. Sometimes this approach is also referred to as intellectual. The implementation of this approach demands a large amount of training data that contain both raw measurements of smartphone sensors and reference solutions of PDR sub-problems.

The main contribution of the work is the large and diversified dataset that may be used as such training data. It contains about 1200 sets of inertial measurements from smartphone sensors. These measurements are called *core data*. The core data were collected while participants walked along flat closed trajectories during 1–10 min. The core data are accompanied by the high accuracy ground truth data obtained by two foot-mounted inertial measurement units (IMUs) and handled by our algorithms that use the ZUPT-method, joint processing of left and right IMU measurements, and the knowledge on flatness and closure of a trajectory. We call our dataset RuDaCoP, which is the acronym for "Russian Data Collection for Pedestrians".

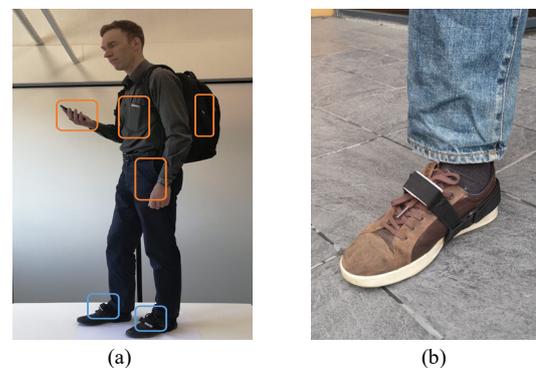

Fig. 1. The example of equipment placement for the dataset collection: (a) four smartphones with arbitrary placement (highlighted in orange) and two inertial measurement units attached to midfeet (highlighted in blue); (b) mounting of an inertial measurement unit

Over one hundred people of both genders and different age from 18 to 60 years old have participated in the collection of experimental data. The experiments vary in participant speed, walking type, footwear, floor type, and placement of smartphones. Note that the presented dataset may be useful not only for machine learning but also for development of more classical PDR algorithms.

Another contribution of our work is the set of the algorithms for the construction of accurate reference data based on dual foot-mounted IMUs and verification of their consistency.

The rest of this paper is organized as follows. Section II contains a review of previously published datasets and their comparison with the dataset presented in this paper. Section III describes how the data collection process was organized in terms of equipment, software, and general logic of experiments. The details of using the dataset along with the reference data are described there as well. Section IV describes the algorithms for processing and verification of the reference data. Section V describes examples of dataset usage. Finally, the conclusions are given in Section VI.

## II. Related Work

Despite the considerable attention to smartphone-based PDR research these days, very few datasets can be used to solve all PDR-related issues. The paper [5], published on 2018, September, directly claims that "no SLE-oriented datasets have been found... none of them [datasets available] are suitable for the task of evaluating SLE methods. The most common limitation is the lack of ground truth information at the required level of accuracy and resolution". Not arguing in general, further, we will mention several notable datasets, emphasizing their differences from the dataset presented in our paper.

Many datasets are dedicated to some specific tasks such as, for example, motion type detection. In these cases, the ground truth does not usually contain full trajectories and its accuracy is often insufficient or unknown. Therefore, the dataset described in [6] is intended for study of motion mode and device mode determination problems. There is no report on accuracy of the ground truth, which is obtained from the smartphone GPS. The dataset described in [7] has relatively small size (100 experiments about 2 min length each, including all kinds of activities such as walking, running, ascending stairs, descending stairs, skipping, standing still). This dataset suits more for training of activity recognition algorithms. The dataset described in [8] is intended for developing an automated fall detection system and contains many such motion patterns as falling, lying, sitting etc. Only a small part of the dataset consists of walking. The ground truth trajectories are absent. The dataset described in [9] was mainly collected by blind volunteers, using a long cane or a guide dog. Since motion patterns of blind and sighted humans differ significantly, this dataset may be used for study of PDR problems for sighted humans with a great care. Then, the ground truth of this dataset contains only heel strike times, segmentation into straight and turn intervals, and annotations of particular events such as opening a door, bumping into an obstacle, stopping momentarily, etc.

Other groups of datasets contain poor variety of trajectories or positions of smartphones. For example, the dataset described in [10] is of interest for PDR development, but it has a lack of diversity of trajectories (all experiments were held with only five different trajectories) and smartphone position (it was always held in a hand in front of a body). The dataset described in [11] contains only straight paths. The ground truth contains only times of walk start and finish, walking/non-walking indication, and a step count. The specifics of the very large database described in [12] is that, firstly, participants walked only along straight paths at varying inclinations, and, secondly, a smartphone was always located in a belt around the waist of a participant.

Some datasets as one described in [13] contain simulated inertial data. This is useful for development of PDR algorithms but obviously simulated data cannot totally substitute real experimental data.

This review increase assurance that our dataset may partly fulfill an existing gap and should be useful for development of PDR algorithms.

## III. Collection process and dataset description

Designing the dataset, we pursued the following goals: accuracy, size, and reliability. The accuracy means that for every measurement we should know exactly where a participant resides, and how he is oriented. The size supposes our attempt to collect as many respondents with different physiological parameters and as many tracks with different parameters as it needs to assume this dataset to be large enough for the training of ML-based algorithms. The reliability marks sufficient knowledge on how exactly the data were collected for every experiment in terms of characteristics of participants and environment as well as actions of participants.

Overall, the collected dataset consists of 1) core data that can be used for the further research on smartphone-based pedestrian navigation, and 2) reference data representing the ground truth. The reference data can be used both as training data for supervised learning techniques and as verification data for classical inertial algorithms that are run on the core data. The core data are without any processing – they are exactly as read from sensors of smartphones through Android OS. The reference data are post-processed to make themselves more precise and to ease further deal with them. The details of post-processing are covered in Section IV. Also for most experiments, information on conditions is available, such as a type of flooring, a type of shoes, anthropometric data of participants (gender, age, weight, height), placement of smartphones (pockets, bags, backpacks, hands, etc.). For some experiments, there are videos of an entire process. We used videos mainly to clarify reasons of rare artefacts in the core and reference data and to exclude the experiments with the artefacts from the dataset.

### A. Participants and trajectories

One of the main challenges for pedestrian navigation algorithms is a necessity of adaptation for physiology of a user and circumstances of using the navigation. That is a reason why we have tried to make a significant diversification for conditions of experiments in such terms as: 1) age of participants (from 18 to 60 years old), 2) gender of participants (about a quarter were females), 3) shoes (there are trajectories in sneakers, gumshoes, business shoes, high-heels, moccasins), 4) different flooring (parquet, asphalt, concrete, linoleum, carpet), 5) placement of smartphones (as mentioned earlier). We asked participants to keep smartphones freely in convenient places, and do not think much about speed or type of walking – just behave naturally.

Still we made some standardization and restriction on trajectories that were walked by participants.

- The trajectories are on flat horizontal surfaces – no stairs or significant changes in landscape heights.

- All trajectories are closed-loop. That means the point of start is equal to the point of finish. The participants were asked to use a paper marker for foot positions, which makes possible to state that the difference in start and finish position is not more than 5 cm for each foot.

- We restricted the participants with walking only (no running/jumping/jibbing or any other strange type of movement).

- We did not restrict the participants on speed of walking.

- Each trajectory is preceded by 10–15 seconds of immobility, and the same pause is in the end of each trajectory.

- The dataset mainly consists of 1 min, 3 min, 5 min, and 10 min trajectories.

### B. Equipment

Each equipment set we used for dataset collection consists of the entities shown in Fig. 2. They included four smartphones for gathering the core data. The smartphones had arbitrary placement in pockets, hands, bags of a participant. Then, the equipment includes two foot-mounted inertial measurement units for reference data collection. These modules were mounted to the upper part of participants' midfoot with rubber bindings, as shown in Fig. 1(b), to keep them immobile with respect to a foot. An example of all hardware placement is shown in Fig. 1(a). All hardware was connected wirelessly by either Bluetooth 3.1 or Bluetooth 4.1 protocols depending on the exact hardware set.

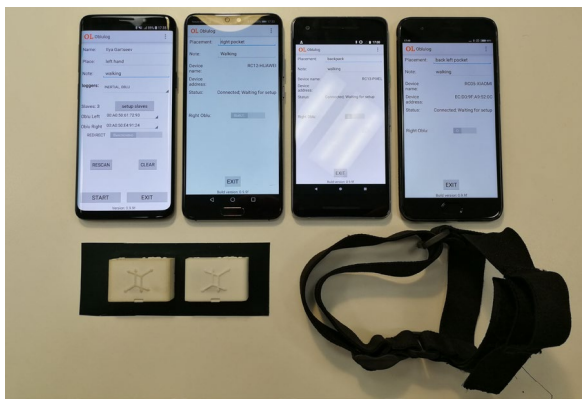

Fig. 2. The equipment set for dataset collection consisted of four smartphones and two IMUs with rubber bindings

The most experiments contain core data collected with four smartphones. There were some rare faults with one–two smartphones during experiments. For such cases, the whole data for such smartphones are absent. Placement of the smartphones is described by key "Placement" in the file *meta.txt* of the dataset. The typical values are "in jeans rear pocket", "in right front pocket of shirt", "in a bag", "in a backpack", "left hand, navigation position", "in swinging hand", etc. One smartphone per experiment always resided in the handheld position (also called "navigation position"), the others were in arbitrary positions chosen by a participant. The smartphones were put into their positions before the beginning of the experiments and the participants were asked not to change the placement of the smartphones during walks.

The dataset was collected by smartphones of different manufacturers to exclude possible problems related to a specific model. The base smartphone set included the following models: 1) Xiaomi MI6; 2) Huawei Nova2 PIC-LX9; 3) Sony Xperia XZ1 G8342; 4) Huawei Honor 9 STF-L09; 5) Huawei P10 VTR-L29; 6) Huawei P20 EML-L29; 7) Google Pixel 2; 8) Samsung Galaxy S9 SM-960F/DS. Also, for some experiments, the following models were used: 1) Google Pixel; 2) Samsung Galaxy S8; 3) OnePlus A5000; 4) Huawei Honor 8. The file *meta.txt* contains explicit information for the used device.

For researchers, the models of the used sensors would be of interest rather than the models of the smartphones. Depending on a smartphone, for a measurement of specific forces and angular velocities, the following 6DOF accelerometers/gyros with the highlighted output data rates were used: 1) Bosch BMI160 (200 Hz/400 Hz for specific forces; 100 Hz/400 Hz for angular velocities); 2) ICM20690 (200 Hz/500 Hz for specific forces; 100 Hz/500 Hz for angular velocities); 3) LSM6DSL (500 Hz for specific forces; 500 Hz for angular velocities); 4) LSM6DSM (400 Hz/500 Hz for specific forces; 400 Hz/500 Hz for angular velocities); 5) unknown ST-based sensor (200 Hz for specific forces; 200 Hz for angular velocities). An exact model of a sensor for each experiment may be automatically restored from the accompanying documents discussed later in the current section (files *accelerometers_*.csv*, *gyroscope_*.csv*). It should be noticed that for most experiments, the smartphones collect not only raw (uncalibrated) gyroscope measurements but also the additional data that processed by inbuilt smartphone algorithms. The uncalibrated data are marked with the word "uncalibrated" in the file name. The typical inbuilt algorithm performs calibration of the sensors by excluding a bias from the measurements.

The following 3DOF magnetometers were used to measure magnetic field: AK09911, AK09915, AK09916C, and AKM09918 with output data rate in the range 50 Hz-200 Hz. For most experiments, both uncalibrated and calibrated data are presented.

For collecting the reference data we used the modules MIMU22BTPX, MIMU22B9PX, and MIMU22BLPX [13], [14] based on integrated circuits Invensense MPU-9150 and Invensense MPU-9250 [15]. A choice of these devices was done due to their four 9DOF IMU array (accelerometer + gyro + magnetometer), placement and orientation of the IMUs to minimize systematic errors, Bluetooth v4.1 interface for MIMU22BLPX, good battery, suitable size and weight (42.2x27.9x17.0 mm; 20 g for MIMU22BLPX). In order to gather the data from the devices, Bluetooth 4.1 protocol was used for MIMU22BLPX modules, and Bluetooth 3.1 protocol was used for MIMU22BTPX, MIMU22B9PX modules. That was enough to get the data with the sampling rate of 125 Hz. The data were logged in one of the smartphones.

### C. Data collection and synchronization

We have developed our own application for the data collection. Fig. 3 shows the logic and the entities of the data collection process.

One of the smartphones (further referred to as the master) contains the interface for control of the process. It starts and stops the data collection by sending the corresponding commands to other smartphones (further referred to as the slaves). The master also initiates and terminates the data gathering process from the foot-mounted IMUs. During an experiment, the data from each smartphone are collected inside the smartphone, the data from IMUs are collected inside the master. The combining and time alignment of the data are performed later off-line.

For each experiment presented in the dataset, all data are saved in the set of folders. One folder corresponds to the reference data of the experiment and other folders correspond to the smartphones used in the experiment. The folders have names appeared as 2018-08-27_18-20-06.730_ 358351080456283. The folder name is constructed as a concatenation of 1) experiment date in the format: YYYY-MM-DD, 2) experiment start time in the format HH-MM-SS.ms, 3) and, finally, an identifier. If the folder contains reference data then the identifier is the word "reference". If the folder contains core data then the identifier is a number uniquely matching to the smartphone that has measured the core data.

One of the main challenges accompanying massive data collection relates to time synchronization of different devices. In our case, this turns into time synchronization of 1) sensor data from all smartphones, 2) sensor data from two reference IMUs on feet, 3) reference trajectories obtained by post-processing algorithms. All smartphones independently logged their own measurements, and each measurement is marked by an Android OS time related to the moment when the measurement comes from a hardware. The accuracy of the interval between two consequent timestamps might vary a little. However, it can be easily restored using the sensor output data rate. We assumed that absolute time in all smartphones are well calibrated.

In order to synchronize timestamps between smartphones, the master translates its start time of the experiment to the slaves along with the command to start logging. Each slave writes to the logs both its own start time and the start time of the master. This information could be found in the file *meta.txt* under the following keys: 1) *MasterSendStartRealtime* is the time of the master when the START command is sent to the slaves; 2) *SlaveReceiveStartRealtime* is the time of the slave when the START command is received from the master. The key *SlaveReceiveStartRealtime* is present only in *meta.txt* of slave smartphones. Thus, to get the common time grid of the experiment one should: 1) subtract *MasterSendStartRealtime* from all timestamps for the master; 2) subtract *SlaveReceiveStartRealtime* from all timestamps for the slave. There remains a slight difference between time scales of smartphones due to delays in a wireless channel but empirical estimation shows that it is under 1 ms for most cases.

The synchronization for the reference data is done by the post-processing algorithm, so the timestamps in the reference data already correspond to the aforementioned common time grid.

### D. Core data description

Each folder with core data contains the files described in Table I. The asterisk denotes the specific natural number. This number might be different for each smartphone participating in the experiment.

TABLE I. DESCRIPTION OF THE FOLDERS WITH CORE DATA

| Filename | Content | Details |
|---|---|---|
| accelerometer_*.csv | Specific forces data | Timestamps in ms; Specific force along 3 axes (including gravity) in $m/s^2$ |
| gyroscope_*.csv | Angular velocity data (calibrated) | Timestamps in ms; Rate of rotation around 3 axes in rad/s |
| gyroscope_uncalibrated_*.csv | Angular velocity data (uncalibrated) | Timestamps in ms; Rate of rotation (without drift compensation) around 3 axes in rad/s; Estimated drift of angular velocities in rad/s |
| magnetic_field_*.csv | Magnetic field data (calibrated) | Timestamps in ms; Geomagnetic field strength along the 3 axes in µT |
| magnetic_field_uncalibrated_*.csv | Magnetic field data (uncalibrated) | Timestamps in ms; Geomagnetic field strength (without hard iron calibration) along the 3 axes in µT; Iron bias estimation along the 3 axes in µT |
| meta.txt | Auxiliary data about the experiment | Organized as dictionary with some technical details on an experiment. The key 'Placement' refers to text description of smartphone position. The key 'Note' refers to any additional info that participant would like to mention |

### E. Reference data description

For each experiment, there are three CSV-files with reference data. The file *Trajectory.csv* contains three smoothed trajectories: left foot, right foot, and participant's center of gravity. Also it contains Boolean indicators of motionless for the left and right feet. Each (n+1)-st row in the file *Trajectory.csv* corresponds to the moment $t_n$ of a single IMU measurement. Recall that these measurements have frequency of 125 Hz. Content of this file is described in Table II.

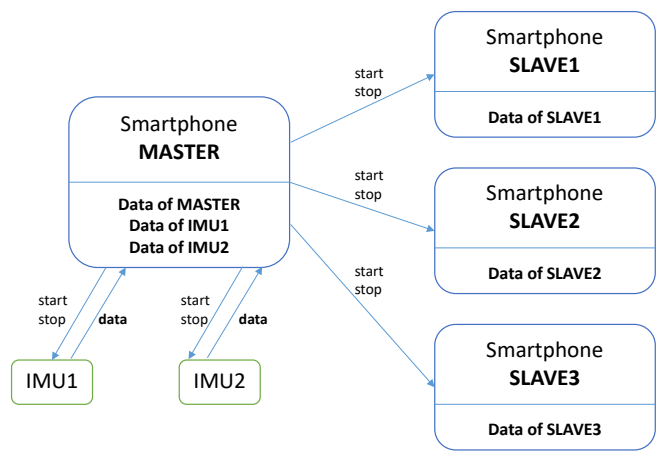

Fig. 3. The elements of the dataset collection process and their relations

TABLE II.     CONTENT OF TRAJECTORY.CSV

| Column number | Key | Value | Dimension |
|---|---|---|---|
| 1 | t[s] | $t_n$ | seconds |
| 2 | x_left[m] | $(\mathbf{p}_n^L)_x$ | meters |
| 3 | y_left[m] | $(\mathbf{p}_n^L)_y$ | meters |
| 4 | left_stationary | $S_n^L$ | Boolean |
| 5 | x_right[m] | $(\mathbf{p}_n^R)_x$ | meters |
| 6 | y_right[m] | $(\mathbf{p}_n^R)_y$ | meters |
| 7 | right_stationary | $S_n^R$ | Boolean |
| 8 | x_avg[m] | $(\mathbf{r}_n)_x$ | meters |
| 9 | y_avg[m] | $(\mathbf{r}_n)_y$ | meters |

The notations used in the table and later are introduced as follows:

- XYU is a right-handed Cartesian coordinate system such that the third axis is directed upwards, parallel to the gravity vector. The first and the second axes have arbitrary placement in the horizontal plane.
- $a_x$, $a_y$, and $a_u$ are coordinates of any 3D vector $\mathbf{a}$ in the coordinate system XYU.
- $\{t_n\}_{n\in\{1,...,N\}}$ are all time moments when the IMUs on the left and right feet simultaneously get values from their accelerometers and gyroscopes.

For every $n \in \{1, ..., N\}$,

- $\mathbf{f}_n^L$ is a 3D vector of an accelerometer measurement of the left foot IMU at the moment $t_n$ (hereinafter the upper index '$L$' denotes the left foot).
- $\mathbf{w}_n^L$ is a 3D vector of a gyroscope measurement of the left foot IMU at the moment $t_n$.
- $S_n^L$ is a Boolean value that indicates whether the left foot is motionless on a floor at the moment $t_n$ ('1' denotes motionless state).
- $\mathbf{p}_n^L$ is a 3D vector of a left foot position in the coordinate system XYU at the moment $t_n$.
- $\mathbf{v}_n^L$ is a 3D vector of a left foot velocity in the coordinate system XYU at the moment $t_n$.
- $\mathbf{r}_n$ is a 3D vector of the position of participant's center of gravity in the coordinate system XYU at the moment $t_n$.

We use the similar notation replacing the upper index '$L$' by '$R$' for the right foot. We omit these indices when specification of a foot is not needed.

The files *Left_steps.csv* and *Right_steps.csv* contain trajectories of both feet. The trajectories are represented by footstep lengths and directions. Each $(k + 1)$-st row in the file *Left_steps.csv* corresponds to the moment $\tau_k^L$ of a step start. Content of this file is described in Table III. Content of *Right_steps.csv* is similar.

TABLE III.     CONTENT OF LEFT_STEPS.CSV

| Column number | Key | Value | Dimension |
|---|---|---|---|
| 1 | t[s] | $\tau_k^L$ | seconds |
| 2 | length[m] | $\lambda_k^L$ | meters |
| 3 | theta[rad] | $\theta_k^L$ | radians |

The notations used in the Table III and later are introduced as follows.

Let $\{\tau_k^L\}_k \subset \{t_n\}_{n\in\{1,...,N\}}$ be all time moments when the left foot starts to move and thus begins a step. We may use this set of moments as a step detector and to count steps. For every $k$,

- $\nu_k^L$ is a unique solution of the equation $t_n = \tau_k^L$ for the integer $n$; then $t_{\nu_k^L} = \tau_k^L$.
- $\mathbf{d}_k^L \stackrel{\text{def}}{=} \mathbf{p}_{\nu_{k+1}^L}^L - \mathbf{p}_{\nu_k^L}^L$ is a 3D vector of full shift of the left foot in the coordinate system XYU during $k$-th step.
- $\lambda_k^L \stackrel{\text{def}}{=} \sqrt{(\mathbf{d}_k^L)_x^2 + (\mathbf{d}_k^L)_y^2}$ is a *horizontal length* of $k$-th step of the left foot.
- $\theta_k^L$ is a *heading angle* of $k$-th step of the left foot; this means that $\theta_k^L \in \text{Arg}\big((\mathbf{d}_k^L)_x + i(\mathbf{d}_k^L)_y\big)$ and $|\theta_{k+1}^L - \theta_k^L| < \pi$.

All results for the reference data are obtained by the post-processing algorithm using foot-mounted IMU measurements and information about flatness and closeness of a trajectory. The algorithm is described in Section IV.

## IV. THE ALGORITHMS FOR REFERENCE DATA PROCESSING

### A. Reconstruction of trajectories

The dynamic system that describes the motion of a foot-mounted IMU has the following form:

$$\begin{cases} \mathbf{p}_n = \mathbf{p}_{n-1} + \mathbf{v}_{n-1} dt_n, \\ \mathbf{v}_n = \mathbf{v}_{n-1} + (\widetilde{\mathbf{C}}_n^T \bar{\mathbf{f}}_n + \mathbf{g}) dt_n, \\ \widetilde{\mathbf{C}}_n = \mathbf{V}(\widetilde{\mathbf{w}}_n dt_n)\, \widetilde{\mathbf{C}}_{n-1}, \end{cases} \quad (1)$$

where $dt_n \stackrel{\text{def}}{=} t_n - t_{n-1}$, $\mathbf{g}$ is the gravity vector in the frame XYU; $\widetilde{\mathbf{C}}_n \in \mathbb{R}^{3\times3}$ is the orientation matrix of the body frame relative to the frame XYU at the moment $t_n$, $\bar{\mathbf{f}}_n = \mathbf{f}_n + \delta \mathbf{f}_n$ and $\widetilde{\mathbf{w}}_n = \mathbf{w}_n + \delta \mathbf{w}_n$ are the exact values of specific force and angular velocity, $\mathbf{V}(\widetilde{\mathbf{w}}_n dt_n)$ is the matrix of rotation by the vector-angle $\widetilde{\mathbf{w}}_n dt_n$.

Since system (1) is nonlinear, one has to linearize it to apply classical approaches. Let $\bar{\mathbf{C}} = \bar{\mathbf{C}}(t)$ be the exact orientation matrix of an IMU and $\mathbf{C} = \mathbf{C}(t)$ be the calculated orientation matrix. Using a small vector of rotation $\boldsymbol{\beta}$, we write:

$$\bar{\mathbf{C}}^T = \big(\mathbf{I}_{3\times3} + \widehat{\boldsymbol{\beta}}\big)\mathbf{C}^T, \quad (2)$$

where $\mathbf{I}_{3\times3} \in \mathbb{R}^{3\times3}$ is the identity matrix and $\widehat{\boldsymbol{\beta}}$ is a skew-symmetric matrix:

$$\widehat{\boldsymbol{\beta}} = \begin{bmatrix} 0 & \beta_3 & -\beta_2 \\ -\beta_3 & 0 & \beta_1 \\ \beta_2 & -\beta_1 & 0 \end{bmatrix} \in \mathbb{R}^{3\times3}.$$

It is possible to consider $\mathbf{x}_n = [\mathbf{p}_n \ \mathbf{v}_n \ \boldsymbol{\beta}_n]^T$ as a state vector and thus consider the following linear system:

$$\mathbf{x}_n = \mathbf{F}_n \mathbf{x}_{n-1} + \mathbf{G}_n \begin{bmatrix} \delta \mathbf{f}_n \\ \delta \mathbf{w}_n \end{bmatrix} + \mathbf{L}_n, \qquad (3)$$

$$\mathbf{F}_n = \begin{bmatrix} \mathbf{I}_{3\times3} & \mathbf{I}_{3\times3}dt_n & \mathbf{0}_{3\times3} \\ \mathbf{0}_{3\times3} & \mathbf{I}_{3\times3} & -\overline{\mathbf{C}_n^T \mathbf{f}_n} \, dt_n \\ \mathbf{0}_{3\times3} & \mathbf{0}_{3\times3} & \mathbf{I}_{3\times3} \end{bmatrix},$$

$$\mathbf{G}_n = \begin{bmatrix} \mathbf{0}_{3\times3} & \mathbf{0}_{3\times3} \\ \mathbf{C}_n^T dt_n & \mathbf{0}_{3\times3} \\ \mathbf{0}_{3\times3} & -\mathbf{C}_n^T dt_n \end{bmatrix}, \quad \mathbf{L}_n = \begin{bmatrix} \mathbf{0}_{3\times3} \\ (\mathbf{C}_n^T \mathbf{f}_n + \mathbf{g})dt_n \\ \mathbf{0}_{3\times3} \end{bmatrix},$$

where $\mathbf{0}_{3\times3} \in \mathbb{R}^{3\times3}$ is a matrix of zero elements.

In order to reconstruct the presented trajectories, we used the algorithms of a Kalman filter type. The algorithms are described in details in [16].

We assume that the foot is on the ground at the time moment $t_n$ if $T(\mathbf{f}_{W_n}, \mathbf{w}_{W_n}) < \gamma$ where $T$ is the function introduced in [17], $W_n = \{n - h, n - h + 1, \ldots, n + h\}$ is a set of indices, $h$ is a positive integer, $\gamma$ is a fixed constant. Then it is possible to use the condition $\mathbf{v}_n = \mathbf{0}$. Since the trajectories are closed-loop, the information about the final position $\mathbf{p}_{final} = \mathbf{0}$ can also be used (without loss of generality we consider it to be zero).

Since the calculated trajectories become discontinuous at the moments of Kalman filter correction, and since the incoming observations do not affect the already constructed parts of the trajectories, it was proposed to smooth the entire trajectory using the Rauch-Tung-Striebel (RTS) filter [18]. However, the angle values are corrected at the forward Kalman filter stage to avoid discrepancy between equations (1) and linear approximation (3).

The corresponding pseudo code is presented in Fig. 4. The following notations are used: $\mathbf{Q} \in \mathbb{R}^{6\times6}$ is a covariance matrix of the measurement noise, $\mathbf{R}' \in \mathbb{R}^{6\times6}$ and $\mathbf{R}'' \in \mathbb{R}^{3\times3}$ are covariance matrices for the noises of the additional conditions $[\mathbf{p}_n \mathbf{v}_n]^T = \mathbf{0}$ and $\mathbf{v}_n = \mathbf{0}$ respectively. The function $f_{mech}$ stands for equations (1).

The initial heading angles cannot be restored without additional information. However, the difference between heading angles of different feet can be determined with the aforementioned algorithm. For this purpose, we minimize the value of DTW-distance [19] between the calculated trajectories using the brute-force search.

After that, we use the algorithm for construction of two "fused" trajectories as described in [16], which takes into account the information about position of the other foot in the end of each step. In this case the foot that makes first step is determined using the function $T(\cdot)$, and then the trajectories of both feet are calculated side by side in a similar way. The RTS-smoothing is also applied.

To merge two obtained trajectories into one, we interpolate the trajectories to the common time grid $\{t_n\}_{n\in\{1,\ldots,N\}}$. These interpolated trajectories are estimations for the values $\{\mathbf{p}_n^L\}_{n\in\{1,\ldots,N\}}$ and $\{\mathbf{p}_n^R\}_{n\in\{1,\ldots,N\}}$. Then we assign the position of participant's center of gravity by the formula: $\mathbf{r}_n = (\mathbf{p}_n^L + \mathbf{p}_n^R)/2$.

---

**Initialization:** $\mathbf{x}_0 \leftarrow E[\mathbf{x}_0]$, $\delta\mathbf{x}_0 \leftarrow \mathbf{0}$, $\mathbf{P}_0 \leftarrow \text{cov}(\mathbf{x}_0)$
/* Forward Kalman filter stage */
**for** $n = 2$ to $N$ **do**
  $\mathbf{x}_n \leftarrow f_{mech}(\mathbf{x}_{n-1}, \mathbf{f}_n, \mathbf{w}_n)$
  $\delta\mathbf{x}_{n|n-1} \leftarrow \mathbf{F}_n \delta\mathbf{x}_{n-1|n-1}$
  $\mathbf{P}_{n|n-1} \leftarrow \mathbf{F}_n \mathbf{P}_{n-1|n-1} \mathbf{F}_n^T + \mathbf{G}_n \mathbf{Q} \mathbf{G}_n^T$
  **if** $T(\mathbf{f}_{W_n}, \mathbf{w}_{W_n}) < \gamma$ **then**
    **if** standstill$(n) = $ true **then**
      $\mathbf{H} \leftarrow \begin{bmatrix} \mathbf{I}_{3\times3} & \mathbf{0}_{3\times3} & \mathbf{0}_{3\times3} \\ \mathbf{0}_{3\times3} & \mathbf{I}_{3\times3} & \mathbf{0}_{3\times3} \end{bmatrix}$
      $\mathbf{K}_n \leftarrow \mathbf{P}_{n|n-1} \mathbf{H}^T \left( \mathbf{H} \mathbf{P}_{n|n-1} \mathbf{H}^T + \mathbf{R}' \right)^{-1}$
      $\delta\mathbf{x}_{n|n} \leftarrow \delta\mathbf{x}_{n|n-1} - \mathbf{K}_n \begin{bmatrix} \delta\mathbf{p}_{n|n-1} - \mathbf{p}_n \\ \delta\mathbf{v}_{n|n-1} - \mathbf{v}_n \end{bmatrix}$
    **else**
      $\mathbf{H} \leftarrow \begin{bmatrix} \mathbf{0}_{3\times3} & \mathbf{I}_{3\times3} & \mathbf{0}_{3\times3} \end{bmatrix}$
      $\mathbf{K}_n \leftarrow \mathbf{P}_{n|n-1} \mathbf{H}^T \left( \mathbf{H} \mathbf{P}_{n|n-1} \mathbf{H}^T + \mathbf{R}'' \right)^{-1}$
      $\delta\mathbf{x}_{n|n} \leftarrow \delta\mathbf{x}_{n|n-1} - \mathbf{K}_n (\delta\mathbf{v}_{n|n-1} - \mathbf{v}_n)$
    **end**
    $\mathbf{P}_{n|n} \leftarrow (\mathbf{I}_{9\times9} - \mathbf{K}_n \mathbf{H}) \mathbf{P}_{n|n-1}$
    /* Compensate internal angle states */
    $\mathbf{C}_n \leftarrow (\mathbf{I}_{3\times3} + \hat{\boldsymbol{\beta}}_n) \mathbf{C}_n$
    $\boldsymbol{\beta}_n \leftarrow \mathbf{0}$
  **end**
**end**
/* Smoothing */
**for** $n = N - 1$ to $1$ **do**
  $\mathbf{A}_n \leftarrow \mathbf{P}_{n|n} \mathbf{F}^T \mathbf{P}_{n+1|n}^{-1}$
  $\delta\mathbf{x}_{n|N} \leftarrow \delta\mathbf{x}_{n|n} + \mathbf{A}_n (\delta\mathbf{x}_{n+1|N} - \delta\mathbf{x}_{n+1|n})$
  $\mathbf{P}_{n|N} \leftarrow \mathbf{P}_{n|n} + \mathbf{A}_n (\mathbf{P}_{n+1|N} - \mathbf{P}_{n+1|n}) \mathbf{A}_n^T$
**end**
/* Compensate internal states */
**for** $n = 1$ to $N$ **do**
  $\mathbf{x}_n \leftarrow \mathbf{x}_n + \delta\mathbf{x}_{n|N}$
  $\delta\mathbf{x}_n \leftarrow \mathbf{0}$
**end**

Fig. 4. The trajectory reconstruction algorithm based on foot-mounted IMU measurements

The result of the proposed algorithm for two IMUs is shown in Fig. 5. The "fused" trajectories of the left and right feet are marked in yellow and cyan respectively, red color stands for the trajectory $\{\mathbf{r}_n\}_{n\in\{1,\ldots,N\}}$.

We estimate the quality of algorithms by proximity of feet trajectories that were obtained in same experiments. The values of DTW-distances across all experiments in the dataset are shown in Fig. 6. Moreover, we did not include a trajectory into the dataset if we had considered it suspicious in terms of this metric.

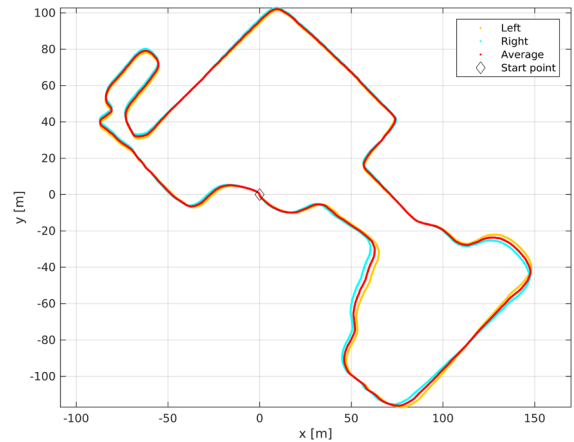

Fig. 5. An example of post-processed reference data including trajectories for right and left feet as well as fused trajectory

## B. Foot motionless detection and step detection

A *foot motionless detection algorithm* constructs the set $\{S_n^L\}_n$ using the sets $\{t_n\}_n$, $\{\mathbf{f}_n^L\}_n$, and $\{\mathbf{w}_n^L\}_n$. The same definition is for the right foot. Let us describe our version of the algorithm omitting the upper indices "$L$" and "$R$".

Let us consider an integer $n > 1$. Let $\mathbf{C}_{n-1}$ be an estimation for the orientation matrix at the moment $t_{n-1}$. Suppose that we can construct it using the previous IMU measurements $\{\mathbf{f}_m\}_{m \in \{1,...,n-1\}}$ and $\{\mathbf{w}_m\}_{m \in \{1,...,n-1\}}$. We use the following algorithm of foot motionless detection:

$$S_n^L = \|\mathbf{w}_n\| \leq \varepsilon \quad \& \quad \left\| \mathbf{C}_{n-1}^T \mathbf{V}\left(-\tfrac{1}{2}\mathbf{w}_n dt_n\right)\mathbf{f}_n + \mathbf{g} \right\| \leq \alpha \|\mathbf{g}\|,$$

where $\varepsilon = 0.5$ rad/s, $\alpha = 0.25$. These parameters may be slightly varied. First operand in the relation for $S_n^L$ implies that the angular velocity $\mathbf{w}_n$ is small; second operand implies that the accelerometer measurement $\mathbf{f}_n$ is sufficiently close to vertical direction.

Sometimes, it is more convenient to use a discrete *step detector* rather than periods of foot motionless. Therefore, we developed the algorithm that constructs the set $\{\tau_k^L\}_k$ from the sets $\{t_n\}_n$ and $\{S_n^L\}_n$ (and the same for the right foot). Its pseudo code (for the left foot) is represented in Fig. 7. The parameter $minFlightTime$ may be slightly varied. The algorithm seeks every motion start ($S_n^L = 0$ & $S_{n-1}^L = 1$) and assigns $\tau_k^L \leftarrow t_{n-1}$ if the foot keeps moving during next $minFlightTime$ seconds. For the right foot the algorithm is same.

## V. THE DATASET APPLICATION EXAMPLES

In order to show the dataset application, we have developed the algorithm of step period estimation. We applied this algorithm to the core data of the random experiment from the dataset. The goal was to compare the result of the algorithm and the step periods of the reference data.

Fig. 8 shows the comparison between the step durations got from the reference data and step durations that were calculated based on smartphone data. It worth noting that smartphones had different placement but all of them were the part of the same experiment.

The reference data on the step duration are taken from the files *Left_steps.csv* and *Right_steps.csv* as the differences

```
minFlightTime ← 0.2[s]
flightTime ← 0
nLast ← 1
k ← 0
for n = 2 to N do
    if S_n^L = 1 then
        if flightTime ≥ minFlightTime then
            flightTime ← 0
            k ← k + 1
            τ_k^L ← t_nLast
        end
    else
        if S_{n-1}^L = 1 then
            flightTime ← 0
            nLast ← n − 1
        end
        flightTime ← flightTime + (t_n − t_{n-1})
    end
end
k ← k + 1
if S_N^L = 1 then
    τ_k^L ← t_N
else
    τ_k^L ← t_nLast
end
```

Fig. 7. The step detector algorithm

between each consecutive moments in the union $\{\tau_k^L\}_k \cup \{\tau_l^R\}_l$ of all moments of beginnings of both left and right steps. The red points in Fig. 8 depict the duration between steps detected by our smartphone step detector using the core data. As it goes from the figures, the estimation of average step duration for all smartphones have good fit to the reference data. However, the smartphones that located in the front and back pockets of the pants have significant noises, sometimes damped. The estimation of step duration from the measurements of the smartphone in a hand is less noisy, especially in the first half of the experiment, but has several outliers because of missing steps.

Another example is training a neural network for pedestrian speed prediction. We utilize the convolutional neural network. The network takes raw smartphone measurements (accelerometers and gyroscopes) from the last 10s as an input and outputs current predicted value of speed. The predicted speed value is produced for every input sample

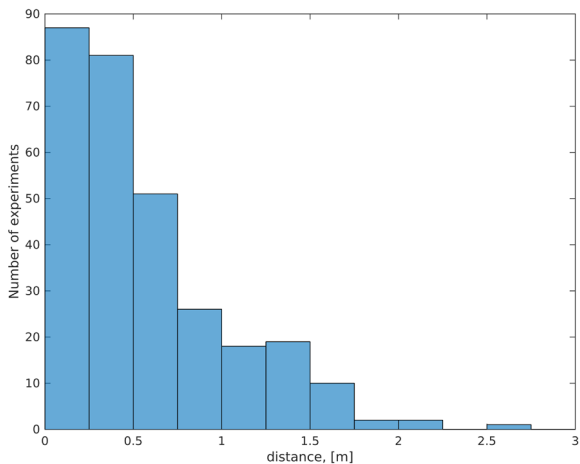

Fig. 6. Distribution of the DTW-distances for all experiments in the dataset

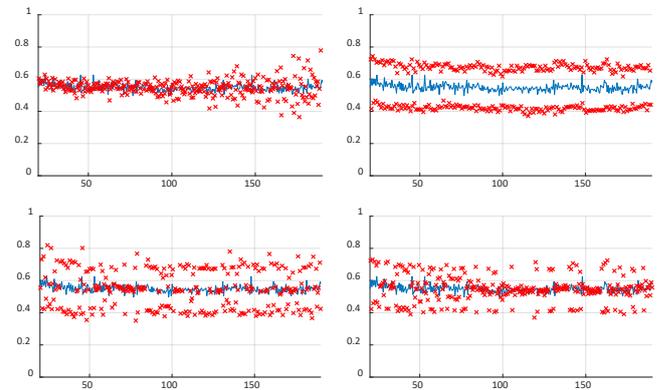

Fig. 8. The duration of the steps detected by smartphones (in red) in comparison with reference data (in blue): (upper-left) a smartphone is in the left hand; (upper-right) a smartphone is in a front right pocket of jeans; (lower-left) a smartphone is in a front left pocket of jeans; (lower-right) a smartphone is in a back right pocket of jeans. For all graphs, x-axis corresponds to experiment time in seconds, y-axis corresponds to duration of steps in seconds

at 100Hz; one can treat it as a continuous version of a step length estimator. A typical output for one experiment is shown in Fig. 9. Average ratio of distance error and true travelled distance is about 3%.

## VI. Conclusions

The paper presents the large and diverse dataset for development of smartphone-based pedestrian navigation algorithms. The logic of experiments, the data collection process, the hardware used, the dataset structure, and the algorithms of computing the ground truth with its verification are described in details. The dataset may be used for training of learning-based pedestrian navigation algorithms and for development of classical PDR algorithms. The discussed examples of the dataset usage include: 1) development of the step detector and its comparison with a ground truth and 2) development of pedestrian speed prediction using deep learning techniques.

The authors hope that this dataset may be useful for researchers who have different backgrounds and who work in the field of pedestrian navigation. It might be useful for the researchers who would like to verify classical navigation approaches to smartphone navigation as well as for the researchers who would like to implement learning-based approaches for the navigation problems.

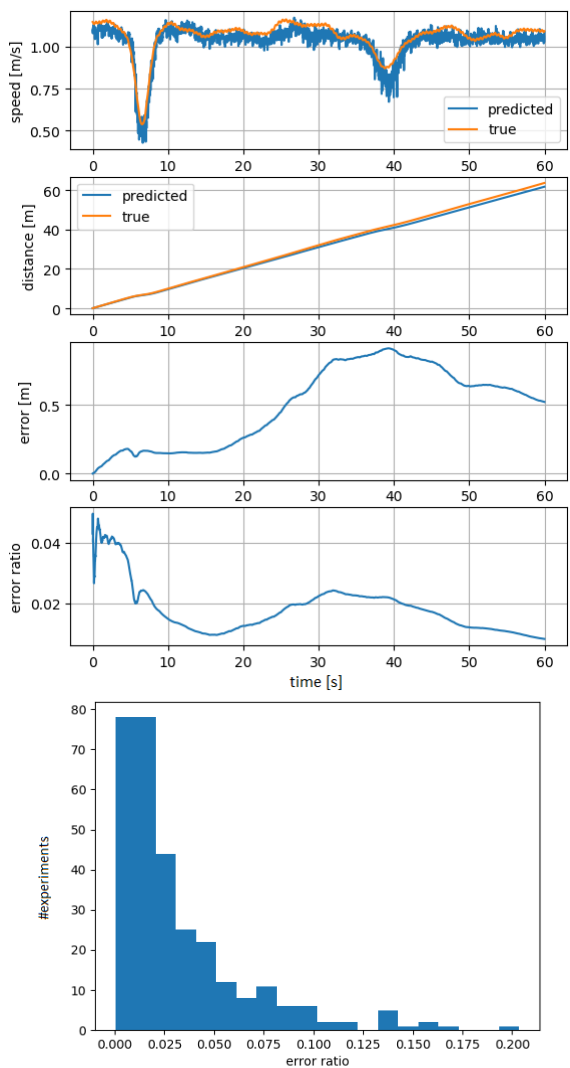

Fig. 9. Sample output of pedestrian speed prediction neural network (upper four graphs), histogram of error ratio (lowermost)

## Acknowledgment

The authors thank Ivan Grishov, Pavel Karpyshev, Dr. Dmitry Tsetserukou, Dr. Mikhail Popelensky, Dr. Andrey Golovan, and Dr. Pavel Tripolsky for their help with different parts of this work.